\documentclass[aps,prx,reprint,secnumarabic,amssymb,amsmath]{revtex4-1}

\usepackage{graphicx,bm}
\usepackage{systeme,braket}
\usepackage[colorlinks=true,linkcolor=blue]{hyperref}%
\expandafter\ifx\csname package@font\endcsname\relax\else
\expandafter\expandafter
\expandafter\usepackage
\expandafter\expandafter
\expandafter{\csname package@font\endcsname}%
\fi
\begin{document}

\title{Perturbation theory for two-dimensional hydrodynamic plasmons}

\author{Aleksandr S. Petrov}%
\email{aleksandr.petrov@phystech.edu}
\affiliation{Laboratory of 2D Materials' Optoelectronics, Moscow Institute of Physics and Technology, Dolgoprudny 141700,	Russia}%

\author{Dmitry Svintsov}%
\affiliation{Laboratory of 2D Materials' Optoelectronics, Moscow Institute of Physics and Technology, Dolgoprudny 141700,	Russia}%

\begin{abstract}
Perturbation theory is an indispensable tool in quantum mechanics and electrodynamics that handles weak effects on particle motion or fields. However, its extension to plasmons involving complex motion of {\it both} particles and fields remained challenging. We show that this challenge can be mastered if electron motion obeys the laws of hydrodynamics, as recently confirmed in experiments with ultra-clean heterostructures. We present a unified approach to evaluate corrections to plasmon spectra induced by carrier drift, magnetic field, Berry curvature, scattering, and viscosity. As a first application, we study the stability of direct current in confined two-dimensional electron systems against self-excitation of plasmons. We show that arbitrarily weak current in the absence of dissipation is unstable provided the structure lacks mirror symmetry. As a second application, we indicate that in extended periodic systems -- plasmonic crystals -- carrier drift induces anomalous Doppler shift, which can be both below and higher than its value in uniform systems. Finally, we exactly evaluate the effect of Berry curvature on spectra of edge plasmons and demonstrate the non-reciprocity induced by anomalous velocity.

\end{abstract}

\maketitle

\section{Introduction and outline}

In quantum mechanics, there is a limited number of potential landscapes that allow exact solutions for energy spectra and wave functions. Fortunately, weak potentials of arbitrary form can be handled with perturbation theory (PT)\cite{SchrodingerPT}. Only several decades after success in quantum mechanics, PT was formulated for classical electrodynamics~\cite{BetheSchwingerPT,WaldronPT}. Currently, electrodynamic PT represents an indispensable tool for analysis of non-uniform laser cavities and inhomogeneous waveguides~\cite{Collin}.

Simplicity of perturbation theory in electrodynamics stems from the fact that state of the field is characterized by two vectors, ${\bf E}$ and ${\bf H}$. Waves propagating near conductive surfaces -- plasmons -- involve not only oscillations of field but of charge carriers as well. The state of carriers is characterized by distribution function generally having an infinite number of harmonics in momentum space. For this reason, state of plasmon is more complex and formulation of plasmonic PT represents a challenging task. Its solution promises a unified approach for treatment of various perturbations on plasma resonances in metal and semiconductor nanostructures, including magnetic fields, electric currents, electron scattering, and others. Previous attempts to construct PT for plasmonic structures required the synthesis of auxiliary equations of motion for polarization and velocity fields in materials that provide a necessary form of dielectric function~\cite{RamanPlasmonsPT}.

In this paper, we show that formulation of simple PT for plasmon eigen frequencies and field distributions is possible when charge carriers in conductors obey the laws of hydrodynamics. While being a common approximation for analysis of carriers motion for nearly a century~\cite{bloch1933bremsvermogen}, the true hydrodynamic phenomena in solids were demonstrated only recently with advent of high-mobility two-dimensional heterostructures~\cite{deJong1995hydrodynamic,bandurin2016negative,crossno2016observation,kumar2017superballistic}. The reason is that the prequisite of hydrodynamics is the dominance of carrier-carrier momentum-conserving collisions over all other collisions (electron-impurity and electron-phonon)~\cite{gurzhi1968hydrodynamic}. Under these conditions, only three harmonics of distribution function survive, and the state of charge carriers is characterized only by three variables: density, velocity, and temperature. 

Electric charge of electron fluid plays the central role in our theory. It leads to existence of plasma modes setting the largest frequency scale in the problem (compared, e.g., to the plasmon decay time). As a result, the unperturbed dynamic matrix is Hermitean, which simplifies the formulation of PT. Such simplicity is lacking in PT for neutral incompressible fluids which motion is strongly affected by viscous dissipation~\cite{joseph1972bifurcating,kuramoto2003chemical}.

\begin{figure}[t]
\includegraphics[width=\linewidth]{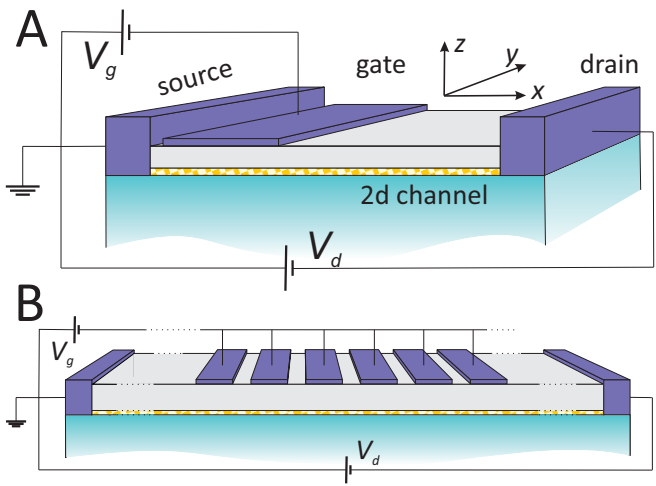}
\caption{\label{Fig1} 
Schematics of 2DES realizations. (A) Bounded 2DES, namely partly-gated field-effect transistor~\cite{petrov2016plasma}. (B) Plasmonic crystal}
\end{figure}

Having constructed the PT for hydrodynamic plasmons, we apply it to physical systems where hydrodynamics was originally observed, namely, to the two-dimensional electronic systems (2DES). As a first example, we study the stability of direct electric current against the excitation of plasmons in confined 2DES [keeping in mind a field-effect transistor (FET) shown in~\ref{Fig1}A]. We succeed to relate the current-induced growth/decay rate of plasmon and its steady-state field distribution in 2DES with arbitrary gating geometry and arbitrary contact boundary conditions. On one hand, this relation aggregates the outcome of preceding excursive studies of plasma instabilities in 2d electron system initiated by Dyakonov and Shur~\cite{dyakonov1993shallow,ryzhii2005transit,petrov2016plasma,sydoruk2010corbino,dyakonov2008boundary}. On the other hand, it shows that structural asymmetry is a necessary and (in the absence of dissipation) sufficient condition for self-excitation of plasmons. This resolves the long-standing experimental puzzle about the relation between structural asymmetry and strength of plasmon-assisted terahertz emission from FETs~\cite{knap2004terahertz,Knap_RT_emission,ElFatimy2010algan}.

The second example is devoted to drift effects in plasmonic crystals (Fig.~\ref{Fig1}B). Existing theories~\cite{chaplik1985absorption,mikhailov1998plasma,kachorovskii2012current} predict the instability excitation only at high velocities (higher than plasma wave velocity) in these structures. We demonstrate that low-drift instabilities are indeed prohibited due to energy conservation (Sec.~IV); still, plasmonic crystals can be used in resonant photodetection providing higher-than Doppler frequency shift in a certain range of parameters. 


The third example addresses the nature of edge modes in anomalous Hall materials hosting 'anomalous' carrier velocity~\cite{xiao2010berry}. We find that this velocity causes non-reciprocity of the system in the \textit{absence of external magnetic field}, introducing the frequency splitting of right- and left-propagating edge modes. The same qualitative result was obtained earlier~\cite{song2016chiral,zhang2018chiral} in restrictive quasi-local electrostatics~\cite{fetter1985edge}; PT allows us to refrain from this approximation and correctly specify the splitting magnitude.

\section{Perturbation theory for electron hydrodynamics}

The confinement of a 2DES to a characteristic length $L$ leads to the emergence of collective modes (plasma modes) with frequencies $\omega_{p} \sim [n_0 e^2 / m \varepsilon L]^{1/2}$, where $e>0$ is the elementary charge, $n_0$ is the electron density, $m$ is the effective mass, and $\varepsilon$ is the background dielectric constant. These modes have been extensively studied since the pioneering works of Stern~\cite{Stern_plasmons} and Chaplik~\cite{chaplik1972possible} and host a variety of phenomena when exposed to external influence, e.g. magnetic field~\cite{Mast_magnetoplasmons} or carrier drift~\cite{Raman_drifting_plasmons}. Interestingly, many of these effects can be incorporated inside a single theoretical shell of plasmonic perturbation theory (PT), which we develop in this Section. 

The formulation of plasmonic PT is possible if oscillating charge carriers obeys the laws of hydrodynamics. Hydrodynamic approach grants a straightforward formulation of operator eigenvalue problem on 2d plasmon eigenfrequencies, allowing the isolation of drift, magnetic field, viscosity and pressure operators --- which is unachievable in other transport regimes. This property enables us to construct an analogy of quantum-mechanical perturbation theory with respect to these operators; such treatment is possible when the corrections to the eigenfrequency are small as compared to the frequency itself. That is why the perturbative approach is much simpler in the case of a charged fluid rather than in conventional hydrodynamics~\cite{joseph1972bifurcating}: the former possesses large zero-order frequency $\omega_p$ that generally outweighs the corrections, whereas the latter deals with much less frequencies and is thus laden with dissipation terms in  the zero-order problem.

The formal inequalities for the perturbative treatment to be possible are $\{u_0/L,\omega_c,\tau^{-1}_p,\nu/L^2\}\ll \omega_{p}$, where $\nu$ is the kinematic viscosity and $\omega_c=eB/m$ is the cyclotron frequency ($B$ denotes the magnetic inductance). Taking the realistic parameters $u_0 \simeq 10^5$\,m/s, $L\simeq 1$ $\mu$m, $\tau_p \simeq 10^{-11}$ s$^{-1}$, $B=0.1\,T$ and estimating the viscosity as $\nu\simeq v^2_0 \tau_{ee}/4\simeq 250$ cm$^2$/s~\cite{kumar2017superballistic}, we see that the inequalities are fulfilled for $\omega_p/2\pi \simeq 1$ THz.

After these preliminary remarks we are ready to construct the theory. The set of HD equations for a two-dimensional motion of a charged fluid has the form:
\begin{gather}
\label{eq-continuity-full}
\partial_t \mathcal N + \partial_i J_i = 0;\\
\label{eq-Euler-full}
\partial_t J_i + \partial_j\mathcal P_{ij} = F_i/m -  J_i/\tau_{\rm p},
\end{gather}
where $t$ denotes time, $\mathcal N$ is the electron density, $J_i= \mathcal N \mathcal U_i$ is the current, $\mathcal U$ is the drift velocity, $F$ is the Lorentz force, and $\mathcal P_{ij}$ is the stress tensor of the electron fluid (we neglect the bulk viscosity and pressure terms, which can be easily restored if needed):
\begin{gather}
	\label{eq-Lorentz}
	\mathbf{F} = e\mathcal N\nabla\varphi + \omega_c [\hat{\mathbf z},\mathcal U]; \\
    \label{eq-Pressure}
    \mathcal P_{ij} = \mathcal N \mathcal U_i \mathcal U_j - \frac{\eta}{m}\left(\partial_j \mathcal U_i + \partial_i \mathcal U_j - \delta_{ij}\partial_k\mathcal U_k\right),
\end{gather}
where $\hat{\mathbf z}$ is the unitary vector in the direction perpendicular to 2DEG (see Fig.~\ref{Fig1}), $\eta$ denotes dynamic viscosity and  $\delta_{ij}$ is the Kronecker delta, $\{i,j,k\}=1,2$. The set (\ref{eq-continuity-full})-(\ref{eq-Pressure}) is completed by the expression for the electric potential $\varphi$ determined by the 2DEG surroundings (consider Fig.~\ref{Fig1}) through the electrostatic Green function $G(\mathbf r,\mathbf r^\prime)$:
\begin{equation}
    \label{eq-Poisson-full}
\varphi(\mathbf r) = \varphi_{ext}(\mathbf r) - e\,\mathcal G[\mathcal N],
\end{equation}
where $\mathcal G[f]=\int {\mathrm d^2\mathbf r^\prime G(\mathbf r,\mathbf r^\prime)f(\mathbf r^\prime)}$ is the self-consistent field, the $\mathbf r$-vector lies in the 2DEG plane, the contribution $\varphi_{ext}(\mathbf r)$ is fixed at the contacts by the voltage source and the integration is performed over the whole 2DEG.

The following analysis is based on linearization $\mathcal N(\mathbf r,t) = n_0(\mathbf r) + n(\mathbf r)e^{-i\Omega t}$, $\mathcal U (\mathbf r,t) = \mathbf u_0(\mathbf r) + \mathbf{u}(\mathbf r)e^{-i\Omega t}$ and reformulation of (\ref{eq-continuity-full})-(\ref{eq-Poisson-full}) as an operator eigenvalue problem:
\begin{equation}
    \label{eq-Matrix}
    (\hat{\Omega}+\hat{V}_{drift}+\hat{V}_{sc}+\hat{V}_{visc}+\hat{V}_{mag})\mathbf{\Phi} = \Omega\mathbf{\Phi}.
\end{equation}
Above, we have introduced the ''three-component wave function'' $\mathbf\Phi = \{ n, \mathbf u\}^{\rm T}$ describing density and velocity variations in plasma modes. The unperturbed motion is described by the 'hydrodynamic' operator:
\begin{equation}
    \label{eq-def-L}
    \hat{\Omega} = 
    -i\begin{pmatrix} 
		0 & \nabla[n_0(\mathbf r)\cdot] \\
      \frac{e^2}{m}\nabla\mathcal G[\cdot] & 0  \\
    \end{pmatrix},
\end{equation}
and we consider the steady carrier drift, magnetic field, scattering, and viscosity as small perturbations given by the operators 
\begin{gather}
    \label{eq-def-Vdrift}
    \hat{V}_{drift} = -i\begin{pmatrix}
    							\nabla(\mathbf{u}_0\cdot) & 0 \\
                                0 & (\cdot,\nabla)\mathbf{u}_0+(\mathbf{u}_0,\nabla)\cdot\\
    							\end{pmatrix}; \\
    \label{eq-def-Vvisc}
    \hat{V}_{visc} =i\begin{pmatrix}
    							0 & 0 \\
                                0 & \eta\Delta\\
    							\end{pmatrix};
     \hat{V}_{sc} =-i\tau_p^{-1}\begin{pmatrix}
    							0 & 0 \\
                                0 & 1\\
    							\end{pmatrix};\\
     \label{eq-def-Vmag}
     \hat{V}_{mag} =-i\omega_c\begin{pmatrix}
    							0 & 0 \\
                                0 & e_{ij}\\
    							\end{pmatrix};
\end{gather}
where $e_{ij}$ is the two-dimensional absolutely antisymmetric tensor, $i,j=1,2$.

At this stage, one might be willing to apply a standard Schroedinger perturbation theory for correction to the eigenfrequency $\delta \Omega_\lambda = \langle \Phi_\lambda | \hat V | \Phi_\lambda \rangle$, where $\lambda$ enumerates the plasmon modes. However, this step is premature until the inner product is specified. Apparently, a standard definition $\langle \Phi_\lambda | \Phi_{\lambda'} \rangle = \int dx[ n_\lambda^* n_{\lambda'} +u_\lambda^* u_{\lambda'}]$ fails: it does not ensure that matrix $\hat \Omega$ is Hermitean. We resolve this issue by reformulating the initial problem: we apply the Hamilton operator $\hat{H}$ of a charged fluid to Eq.(\ref{eq-Matrix}) and obtain a generalized eigenvalue problem:
\begin{gather}
    \label{eq-generalized}
   \hat{H} (\hat{\Omega}+\hat{V}_{drift}+\hat{V}_{sc}+\hat{V}_{visc}+\hat{V}_{mag})\mathbf{\Phi} = \Omega\hat{H}\mathbf{\Phi},\\
   \label{eq-Hamiltonian}
   \hat{H} = 
   \begin{pmatrix}
    e^2/m\,\mathcal G[\cdot] & 0  \\
   0 & \hat{I}_2 n_0(\mathbf r) \\
   \end{pmatrix}.
\end{gather}

Here, the dynamic matrix $\hat{H}\hat{\Omega}$ is Hermitean (i.e. $\bra{\Phi_1}\hat{H}\hat{\Omega}\ket{\Phi_2} = \bra{\Phi_2}\hat{H}\hat{\Omega}\ket{\Phi_1} $) as well as the Hamiltonian $\hat{H}$; this fact can be shown explicitly by evaluating the corresponding matrix elements. Hence, we are now able to apply the standard perturbative expansion that leads to the following expression for the first correction to the  eigenfrequency:
\begin{equation}
    \label{eq-pert}
    \delta\Omega_\lambda = -i\frac{\bra{\Phi_\lambda}\hat{H}(\hat{V}_{drift}+\hat{V}_{sc}+\hat{V}_{visc}+\hat{V}_{mag})\ket{\Phi_\lambda}}{\bra{\Phi_\lambda}\hat{H}\ket{\Phi_\lambda}}.
\end{equation}

We stress that the whole procedure does not require any additional boundary conditions (BC) apart from the two natural ones: (1) the Green function vanishes at the electrodes and (2) $J_y=0$ at the 2DES edges (no carrier leakage). In such a way, our analysis holds for 2DES with \textit{arbitrary} BCs, which are hardly known in real experimental setups.

The constructed perturbation theory is a powerful tool that can be used to uncover the underlying principles of many plasmonic phenomena in 2DES. We apply its formalism to examine the drift-originated plasmonic effects in bounded systems (Sec.~III) and PCs (Sec.~IV), consummating the article with a specific example of Chiral Berry plasmon (Sec.~V).

\section{Current-driven instabilities in bounded systems}

Direct current passing in confined 2DES can supply energy to plasmon modes and lead to their self-excitation (plasma instability). The first example of such conditions was deduced by Dyakonov and Shur who found an instability under source grounded and drain held at fixed current\cite{dyakonov1993shallow}. Later, another geometries and boundary conditions were realized, including loaded drain~\cite{cheremisin1999d}, partly gated FETs~\cite{ryzhii2005transit,petrov2016plasma}, 2DES edges~\cite{dyakonov2008boundary}, and Corbino discs~\cite{sydoruk2010corbino}. This search for instabilities has been excursive and there was no general understanding whether a given FET structure supports an instability or not.

From the prospective of perturbation theory, the operator of electron drift $\hat{V}_{dr}$ in bounded 2DES is non-Hermitean. Therefore, the eigen frequencies of drifting plasmons are generally complex, which implies plasmon amplitude growth/decay with time. The eigenfrequencies may remain real only under specific symmetry constraints which we are to obtain with the developed PT.

We restrict our consideration to one-dimensional oscillations of 2d electrons assuming the 2DES to be uniform in the $y$-direction. Evaluating the matrix elements in Eq.~(\ref{eq-pert}), we find the correction to plasmon frequency in confined 2DES induced by a combined action of direct current, scattering, and viscosity:
\begin{equation}
    \label{eq-main-physical}
    \delta\Omega_\lambda = i\frac{j_0 \left[K_\lambda(0) - K_\lambda(L) \right] - Q_{loss}}{|\Pi_\lambda|},
\end{equation}
where $K(x) = m |u_\lambda(x)|^2/2$ is the local kinetic energy in a plasmon mode,
\begin{equation}
    \Pi = e^2\int\limits_0^L{ \mathrm dx \mathrm dx^\prime n_\lambda^*(x) G(x,x^\prime) n_\lambda(x^\prime)}
\end{equation}
is the potential energy of interacting charge density fluctuations in a 2DES of length $L$, and
\begin{equation}
  Q_{loss} = \frac{1}{2} \int\limits_0^L {\mathrm dx \left\{ \mathrm{Re}\,\sigma |E_\lambda|^2 + \eta |\partial_x u_\lambda|^2\right\}}  +\left.\frac{\eta u^*_\lambda\partial_x u_\lambda}{4}\right|_0^L
\end{equation}
is the energy loss due to viscous friction and Ohm dissipation, where $\sigma = i e^2 n_0/m(\Omega+ i/\tau_{\rm p})$ is the Drude conductivity. The last 'viscous-boundary' term is also dissipative in systems with time-reverse symmetry; if the latter is broken, shear viscosity will contribute to frequency shift as well.

From Eq.~(\ref{eq-main-physical}) we readily observe that plasmon growth rate $\mathrm{Im}\,\delta\Omega_\lambda$ originates from the excess of kinetic energy entering the mode at the source over the energy lost at the drain. A particular example is the DS instability, where $K(0)-K(L)$ is nonzero due to inequivalent source and drain contacts. 

Apart from the energy interpretation, Eq.~(\ref{eq-main-physical}) immediately indicates that only zero-order plasma modes without definite parity can be excited by a weak external drift. Indeed, the even/odd mode profiles require $u^2(L) = u^2(0)$, which forces the gain term in Eq.~(\ref{eq-main-physical}) vanish. Therefore, FETs with mirror symmetry of source and drain do not support unstable modes, but asymmetric structures generally do (in the absence of dissipation). The origin of asymmetry can be arbitrary: either asymmetric placement of gates, or asymmetric loading of source and drain, or non-uniform carrier density in the channel, or all of them.

\begin{figure}[t]
\includegraphics[width=\linewidth]{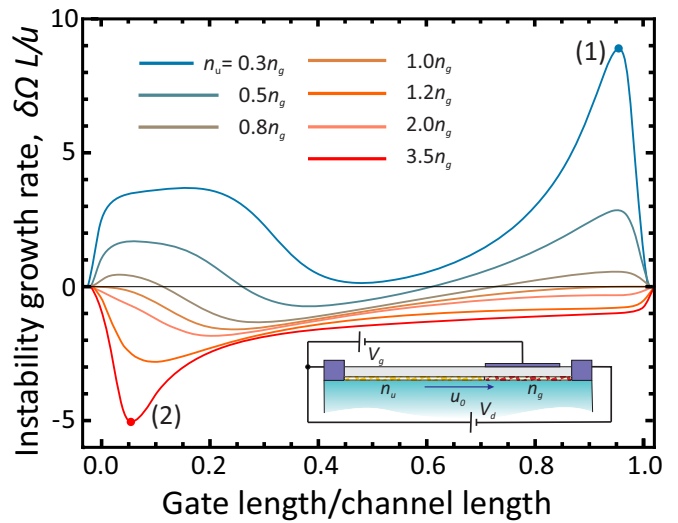}
\caption{\label{Fig2} 
Calculated instability growth rates for the first plasmon mode in a partly-gated FET (shown on the inset) with different gate lengths and carrier densities. The carrier density distribution is taken to be $n(x) = n_1 + (n_2 - n_1)[1+e^{(x-x_0)/w_j}]^{-1}$, $w_j=L/50$ is the junction width. The growth rates are normalized by $u_0/L$, where $u_0$ is the drift velocity at the drain (kept the same for all curves), the density at the drain $n_0$ is also fixed. The instability benefits if the drift is directed from low- into high-density region and is especially pronounced if the low-density region is short and ungated}
\end{figure}

We can go beyond the symmetry constraints and specify the requirements on 2DES aiming to maximize the plasmon growth rate. This is equivalent to maximization of  $u^2(0) - u^2(L)$. The velocity is proportional to electric field and inverse carrier density. Thus, a 2DES with high field and small carrier density at the source and low field and high density at the drain would be most suitable for turbulence. The simplest realization of this scheme is a partly-gated field-effect transistor (FET) with a short ungated depleted region at the source and long, gated and enriched region at the drain (see inset on Fig.~\ref{Fig2}). 


To prove this hypothesis, we develop a toy model for plasma oscillations in a partly-gated FET.  The contacts of such structure are connected to voltage sources, which is a typical experimental situation, and variation of the gate-source bias allows to form a carrier density step in the channel ($n^+-n$ junction). We numerically solve the governing equations in the absence of scattering (Appendix A) and plot in Fig.~\ref{Fig2} the drift-induced corrections to plasmon eigenfrequencies for a set of structures with varying junction location $x_0$ and density modulation factor.  In accordance with our expectations, the highest growth rate (point 1) is achieved for a structure with a short, depleted, ungated source region. If we swap the contacts and thus change the signs of $u_0$ and $\delta\Omega_\lambda$, the magnitude of a new maximum (point 2) would be predictably lower due to the gate screening. In addition, Fig.~\ref{Fig2} shows that the source regions should not be too short, or else the structure would approach the symmetric limits of open ($x_0<0$) or fully-gated ($x_0>1$) FETs with uniform density that are not subject to instabilities. 

Experimental data supports our findings. Thus, the first plasmonic THz sources~\cite{knap2004terahertz} were symmetric and provided broadband radiation only at 4\,K whereas further implication of asymmetric partly-gated FETs~\cite{ElFatimy2010algan} enabled the observation of resonant room-temperature emission due to efficient plasmon-to-drift coupling. This fact apparently shows the need for asymmetry. Moreover, the latter experiment demonstrated that the threshold current is significantly smaller for a depleted source region, which is in agreement with our analysis.

However, this depletion should be kept in proper bounds. Indeed, a sharp density step not only increases  the source field, but also causes highly non-uniform distribution of plasma wave velocity in the channel. This non-uniformity promotes viscous losses, which are proportional to the velocity gradient, and at some degree of asymmetry the viscosity takes over gain. 

\section{Drift-induced phenomena in plasmonic crystals}

It turns out that drift-induced phenomena in PCs is completely different from that in bounded 2DES. The reason is that in periodic systems the drift operator is Hermitean due to translational invariance and thus conserves energy. Still, plasma turbulence in PCs can emerge --- but only at high drift velocities, at least in the depleted regions~\cite{chaplik1985absorption,mikhailov1998plasma,kachorovskii2012current}. In that way, moderate carrier drift in PCs does not affect the stability of plasma modes; instead, it changes their spectrum. 

In order to determine this spectrum change, we apply the constructed perturbation theory with several amendments. To be more specific, after the linearization procedure we exploit a usual Bloch decomposition for the unknown functions and arrive at the same expressions (\ref{eq-def-L})-(\ref{eq-def-Vmag}), but with a modified derivative operator $\partial_x \rightarrow D_x=(\partial_x + iq_B)$ (where $q_B$ is the Bloch vector) and the Green function:
\begin{multline}
    G_{PC}(\mathbf R,\mathbf R')\rightarrow \\ G_{cell}(\mathbf r,\mathbf r^\prime)=\sum_{n=-\infty}^{n=+\infty}G_{PC}(\mathbf r,\mathbf r'+\hat{\mathbf x}nL)e^{iq_B(x'-x)},
\end{multline}
where $\mathbf r$ and $\mathbf r'$ lie within a unit cell of length $L$. The unknowns $n, u,\varphi$ should now be understood as periodic parts of the corresponding Bloch functions. 

These remarks allow us to apply Eq.~(\ref{eq-pert}) to drift-originated phenomena in PCs. We arrive at:
\begin{equation}
\label{eq-corr-pc}
    \frac{\delta \Omega_\lambda}{\Omega_{\lambda}} =j_0 \frac{\int_0^L {\mathrm dx {\rm Im}(mu_\lambda D_x u^*_\lambda)}}{\int_0^L {\mathrm dx {\rm Re}(-en_\lambda \varphi^*_\lambda)}},
\end{equation}
As expected, the correction is real and corresponds to the zero-mode Doppler shift. 

Significant shifts can be useful in resonant photodetection exploiting the plasmonic drag effect~\cite{popov2013terahertz}. Indeed, in a typical PC a normally incident light excites plasma wave packets around $q_B=0$, where the group velocity distribution is symmetrical in the absence of drift. For a sufficient detection, however, one needs a substantial asymmetry in this distribution, which can be introduced via carrier drift. The latter breaks the dispersion curve symmetry, and the greater the Doppler shift, the greater the group velocity difference.

From the first glance, however, PCs may seem not applicable to photodetection: Eq.~(\ref{eq-corr-pc}) implies that $\delta\Omega_\lambda$ turns to zero in the standing-wave limit $q_B=0$ due to ${\rm Im}(mu_\lambda D_x u^*_\lambda) = 0$. Nevertheless, they are. Indeed, the linear correction (\ref{eq-corr-pc}) vanishes in the framework of non-degenerate perturbation theory. But if a PC hosts two close modes, we shall use the degenerate theory and in this case we obtain
\begin{equation}
\label{eq-pert-degenerate}
	\Omega_\pm = \frac12\left\{\Omega_1 + \Omega_2\pm\sqrt{\left(\Omega_1-\Omega_2\right)^2+4|V_{12}|^2}\right\},
\end{equation}
where $V_{12} = \dfrac{\bra{\Phi_1}\hat{H}\hat{V}_{dr}\ket{\Phi_2}}{\bra{\Phi_1}\hat{H}\ket{\Phi_1} \bra{\Phi_2}\hat{H}\ket{\Phi_2} }$.

\begin{figure}[t]
\includegraphics[width=\linewidth]{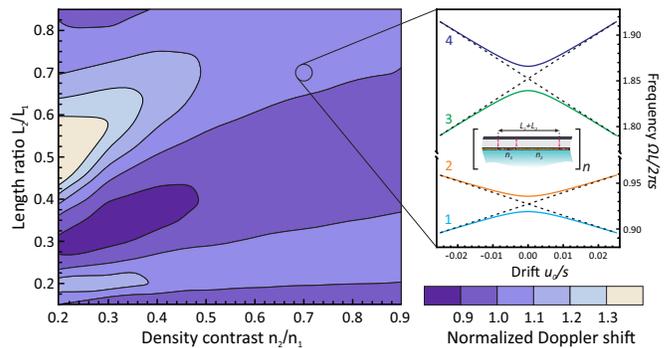}
\caption{\label{Fig3} 
(Left and lower right panels) Normalized Doppler shift for 3rd and 4th modes in fully-gated PCs with a density step inside a unit cell. (Upper right panel) $\Omega(u_0)$ dependencies for a PC with $L_2=0.7L_1$ and $n_2=0.7 n_1$. Their character coincides with the predictions of degenerate perturbation theory Eq.~(\ref{eq-pert-degenerate}). Inset: Scheme of a fully-gated PC}
\end{figure}

Fig.~\ref{Fig3} illustrates our findings. On the left panel we plot the calculated Doppler shifts for 3rd and 4th modes that exist in a range of fully-gated PCs (see inset on the upper right panel) with a density step inside the unit cell. The shifts are normalized by the expected shift value 
\begin{equation}
\dfrac{\Omega}{2\pi sN/L} = \bar{u}_0/N = u_0 \dfrac{L_1/L_2+ n_1/n_2}{NL/L_2},
\end{equation}
where $s$ is plasma wave velocity under the first gate, $L_1$ and $L_2$ are the gate lengths, $L_1+L_2 = L$, $n_2$ and $n_1$ denote carrier densities, and $N$ enumerates the pairs of modes; in our case $N=2$. The upper right panel shows $\Omega(u_0)$ dependencies for the first four modes in a PC with $L_2=0.7L_1$ and $n_2=0.7 n_1$.  In full accordance with Eq.~(\ref{eq-pert-degenerate}), we observe parabolic spectrum at very low velocities that transforms into linear spectrum when the perturbation magnitude exceeds the degeneracy contribution. Hence, the linear part slope is determined by the non-diagonal matrix element and in certain parameter range leads to higher-than Doppler shift.

Actually, there exists another way to break the symmetry constraint ${\rm Im}(mu_\lambda D_x u^*_\lambda) = 0$. Indeed, magnetic field, being included in the $\hat{\Omega}$-operator, efficiently entangles the real and imaginary parts of the zero-order wave functions. Thus, the plasma modes become coupled to drift in the first order (their stability is not affected as as $\hat{V}_{mag}$ is Hermitean). A more detailed discussion of this influence will be given elsewhere.

\section{Chiral Berry plasmons}

Apart from treating the perturbations that directly enter the equations of motion~(\ref{eq-continuity-full})-(\ref{eq-Poisson-full}), our theory also accounts for the boundary condition perturbations. In particular, this property can be useful for the description of Chiral Berry plasmons (CBPs) --- a new type of edge plasmonic states in 2D materials with nonzero Berry flux~\cite{xiao2010berry}. CBPs arise as follows. Nonzero Berry flux provides an 'anomalous' contribution $J_a$ to the total particle current $J_{tot} = J+J_a$, where $J=\mathcal{N}\mathcal{U}$ is the usual current density. The anomalous current does not affect the governing HD equations (due to $\nabla J_a=0$), but changes the boundary condition on the 2DEG edge from the usual $J_{\perp} =0$ to $J_{tot,\perp} = 0$. Song and Rudner~\cite{song2016chiral}, and later Zhang and Vignale~\cite{zhang2018chiral}, found this BC change to split the frequencies of right- and left-propagating plasmons, which were therefore named Chiral Berry plasmons. However, they did that in a restrictive quasi-local electrostatic approximation~\cite{fetter1985edge} that does not, in particular, reproduce the logarithmic divergence of edge magnetoplasmon group velocity at short wave vectors in strong magnetic fields~\cite{volkov1988edge}. This divergence stems from the 2d Coulomb potential divergence, which is violated in the quasi-local approximation. As we demonstrate below, our perturbation theory, which exploits precise electrostatics, reproduces the CBP frequency gap for small Berry flux and corresponding anomalous current. 


We consider a semi-infinite 2DEG with nonzero Berry flux $\mathcal F$ confined to a $\{x>0,y\}$ half-plane and require the potential and electric field continuity at the $x=0$ boundary along with $J_{tot,x}|_{x=+0}= 0$. The latter BC is the only difference that distinguishes the CBP problem from the edge plasmon problem solved by Volkov and Mikhailov~\cite{volkov1988edge}, who obtained the edge plasmon frequency $\Omega_0=\omega_{2d}/\sqrt{1.217}$, $\omega_{2d} = \sqrt{2\pi e^2 n_0|q|/m}$, $q$ is the wave vector in the $y$-direction. Replacing $J$ in Eqs.~(\ref{eq-continuity-full}), (\ref{eq-Euler-full}) with $J_{tot}$ and using $\nabla J_a=0$, we arrive at the following problem:
\begin{equation}
\label{eq-chiral-operator}
(\hat{\Omega}+\hat{V}_a)\Phi = \Omega\Phi,
\end{equation}
where 
\begin{equation}
\label{eq-chiral-v}
\hat{V}_a = 
	\Omega_0 \frac{e\mathcal F}{\hbar n_0}\begin{pmatrix}
  0 & 0 \\
  \nabla_a\mathcal G_q[\cdot] & 0\\
    \end{pmatrix}
\end{equation}
is the 'anomalous' perturbation operator induced by the (dimensionless) Berry flux $\mathcal F$, $\nabla_a=\hat{\mathbf x}iq+\hat{\mathbf y}\partial_x$,  $\mathcal G_q[f]=2\int_0^\infty {\mathrm dx^\prime K_0(q|x-x'|)f(x^\prime)}$, $K_0(x)$ is the modified Bessel function of the second kind, $q$ is the wave vector in the $y$-direction. 

Thus, we transformed the BC perturbation into a more convenient form. Now the formulation of the zero-order problem (\ref{eq-chiral-operator}) coincides with the formulation of edge plasmon problem~\cite{volkov1988edge}. Taking Volkov-Mikhailov mode profiles, we evaluate the Berry-flux-induced correction $\bra{\Phi}\hat{H}\hat{V}_a\ket{\Phi}/\bra{\Phi}\hat{H}\ket{\Phi}$ and calculate the gap between two branches of CBP: 
\begin{equation}
\label{eq-CBP-our}
\hbar\Delta\Omega(q)=3.65 \,\mathcal F e^2|q|
\end{equation}
versus the Song-Rudner result
\begin{equation}
\label{eq-CBP-zhang}
\hbar\Delta\Omega_{SR}(q)=8\sqrt{2}\pi/9\,\mathcal F e^2|q|\approx 3.95\,\mathcal F e^2|q|.
\end{equation}

Thus, our calculations qualitatively approve the approximate Song-Rudner solution and downshift the gap width by 10\%. This change is a minor one and grants all the non-reciprocal implementations of CBPs discussed in~\cite{song2016chiral}.


\section{Discussion and conclusion}


The developed plasmonic PT has the same functionality as PT in quantum mechanics: given the exact solution of unperturbed problem, we can accurately find corrections to eigen-frequency under small perturbations. Unfortunately, exact solutions in 2d plasmonics are unique, among non-trivial cases to mention the edge modes~\cite{volkov1988edge} and plasma oscillations in gated 2DES with infinite conductive walls~\cite{ExactSolution}. At the same time, the unperturbed problem of plasma oscillations due to self-consistent electric field is readily solved with commercial electromagnetic simulators, and the resulting field profiles can be supplied to our perturbation theory. 

The assumption of hydrodynamic transport used in derivation limits the frequencies $\omega$ below the inverse electron-electron scattering time $\tau^{-1}_{ee}$. This may look restrictive as $\tau^{-1}_{ee}$ is order of 1 THz at room temperature~\cite{Crossover} and scales as $T^2$. Most experimental observations of plasmons correspond to the opposite ballistic limit $\omega \tau_{ee} \gg 1$~\cite{Allen,Mast_magnetoplasmons}. However, the difference between predictions of hydrodynamic and ballistic approaches is important only when treating thermal corrections to plasmon velocity and Landau damping~\cite{BGK-collisions}. Therefore, we can speculate that the developed PT would be applicable in the ballistic limit as well.

Another assumption of developed PT is the neglect of retardation effects or, formally, setting the velocity of light $c$ to infinity. This is justified for typical 2d plasmons once their frequency $\omega_0$ lies below the light cone $\omega = c q \sim c/L$~\cite{Retardation}. Renouncement of this assumption immediately leads to radiative plasmon damping and non-Hermiticity of unperturbed problem. Formulation of PT in this case is also possible~\cite{OpenCavities} but requires dealing with diverging fields far away from 2DES.

The presented examples were related to first-order or degenerate perturbation theory. Higher-order corrections can also be derived and are important when first-order effects are absent by symmetry (such as Doppler shift in the center of plasmon Brillouin zone). Another non-trivial application of higher-order corrections is the analysis of weak steady-state plasma turbulence for direct current slightly exceeding the threshold~\cite{landau1944problem,kuramoto2003chemical}. So far, the solution of such problems in 2DES was achieved with numerical simulations~\cite{gabbana2018prospects} or limited to model systems~\cite{dmitriev1996nonlinear}. The general analysis is possible with the developed PT and will be reported elsewhere.

In conclusion, we have developed the perturbation theory for two-dimensional hydrodynamic plasmons and demonstrated its utility on several examples. We have derived a constitutive relation between current-induced growth rate of plasmon and its steady-state field distribution. This expression clarifies the role of structural asymmetry for efficient excitation of plasmons by direct current. In periodic systems -- plasmonic crystals -- the current does not lead to instabilities in the first order, but does induce Doppler shift which can be both above and below the conventional value. Finally, we have shown the capability of PT to handle boundary condition perturbations, and obtained an exact result for left-right edge mode splitting due to Berry curvature.

\section*{Acknowledgements}

The authors thank M. S. Shur and D. Bandurin for valuable discussions. The work was supported by projects 18-37-20058 and 16-37-60110 of the Russian Foundation for Basic Research. A. S. P. acknowledges the support of grant 18-37-00206 of the Russian Foundation for Basic Research and grant 18-1-5-66-1 of the Basis Foundation. 

\appendix

\section{Numerical method}
In order to obtain the results shown in Fig.~\ref{Fig2} we apply a standard spectral numerical method to the system of hydrodynamic equations (\ref{eq-continuity-full}-\ref{eq-Poisson-full}) with Chebyshev polynomials of the first kind $T_i$ taken as the basis functions. To be more concrete, after writing the linearized Eqs. (\ref{eq-continuity-full}), (\ref{eq-Euler-full}) in dimensionless units ($\xi = 2x/L - 1$, $\tilde n = n/n_0(0)$, $\tilde u = u/s(0)$, $\omega = \Omega L/s(0)$ where $s(0)^2 = e^2n_0(0)L/m$) we substitute the Chebyshev expansions in the form 
\begin{equation}
\left\{\tilde n, \tilde u\right\} = \sum_{i=0}^N C_i^{\{n, u\}}T_i(\xi)    
\end{equation}
and project the system on each of the polynomials $T_i(\xi),i = 0..N$. After these manipulations we arrive at the matrix equation:
\begin{equation}
\label{eq-numricalSystem}
    \begin{pmatrix} 
        \hat{M}^{(1)} & \hat{M}^{(2)} \\
        \hat{M}^{(3)} & \hat{M}^{(4)}
    \end{pmatrix}
    \begin{pmatrix} 
        C^{n}_i \\
        C^{u}_i
    \end{pmatrix} = i\omega
    \begin{pmatrix} 
        C^{n}_i \\
        C^{ v}_i
    \end{pmatrix},
\end{equation}
where 
$$\hat{M}_{ij}^{(1)} = \hat{M}_{ij}^{(4)} = t_{ij}\bra{T_j}w(\xi)\partial_\xi \ket{u_0 T_i},$$
$$\hat{M}_{ij}^{(2)} = t_{ij}\bra{T_j}w(\xi)\partial_\xi\ket{n_0 T_i},$$ 
$$\hat{M}_{ij}^{(3)} = t_{ij}\bra{T_j}w(\xi)\partial_\xi\ket{\int_{-1}^1 d\xi'G(\xi,\xi')T_i(\xi')},$$
$$ t_{ij} = \begin{cases} 1/\pi, i=0, \\ 2/\pi, \text{otherwise} \end{cases},$$
and $w(\xi) = (1-\xi^2)^{-1/2}$ is the weight function; $\{i,j\}=0..N$ for all the matrices. 

Implying boundary conditions of fixed charge density at the contacts $\tilde n(-1)=\tilde n(1)=0$, we obtain 
\begin{gather}
 C_N^{n}=-\sum_{i=0}^{N/2-1} C_{2i}^{ n},\\
 C_{N-1}^{n}=-\sum_{i=0}^{N/2-1} C_{2i+1}^{n} , 
\end{gather}
where $N$ is supposed to be even. We use these expressions to eliminate $C_N^{ n}$ and $C_{N-1}^{ n}$ from the system (\ref{eq-numricalSystem}) and, in order to keep the matrix dimensions, truncate the first three matrices.

One may face computational difficulties while evaluating matrix elements $\hat{M}_{ij}^{(3)}$ as the Green function is singular on the diagonal $\xi=\xi'$. To overcome this, we decompose the Green function into singular ($G_s$) and regular ($G_r$) parts:
\begin{equation}
G = (G-G_s)+G_s \equiv G_r+G_s,    
\end{equation}
where 
\begin{equation}
G_{s}=\ln\dfrac{(\xi-\xi')^2}{(\xi+\xi'-2)^2(\xi+\xi'+2)^2}    
\end{equation}
accounts for the singularity provided by the charge itself as well as by the two nearest mirror images in the electrodes. The regular integral is then calculated numerically while the singular one can be significantly simplified via transition to the complex plane.

The Green function of a partly-gated structure is given by
\begin{equation}
    G_{PG}(\xi,\xi^\prime) = \ln\left[\frac{(\alpha-\alpha^{\prime})^2 + (\beta+\beta^{\prime})^2}{(\alpha-\alpha^{\prime})^2 + (\beta-\beta^{\prime})^2}\right],
\end{equation}
where $\alpha^{\prime}+i\beta^{\prime}=z^{\prime}$,
\begin{equation}
 z^{\prime}=
 \cos\psi\sqrt{\tanh^2 \frac{\pi(d + i\xi^{\prime})}{2\,L} + \tan^2\psi}
\end{equation}
and $\psi=\pi L_g/2L$.

The values of the first correction obtained by the described procedure and by the perturbation theory [Eq.(\ref{eq-main-physical})] fully coincide at small drift velocities.

\bibliography{refs,longbibliography}

\end{document}